\documentclass[aps,pra,twocolumn,10pt,superscriptaddress,showpacs,amsmath,amssymb,nofootinbib]{revtex4-1}

\usepackage{graphicx}
\usepackage{hyperref}
\usepackage{array}
\usepackage{color}

\newcommand{\ohm}{$\Omega$}

\newcommand{\framework}{
\begin{figure}[t]
\includegraphics[width=\columnwidth]{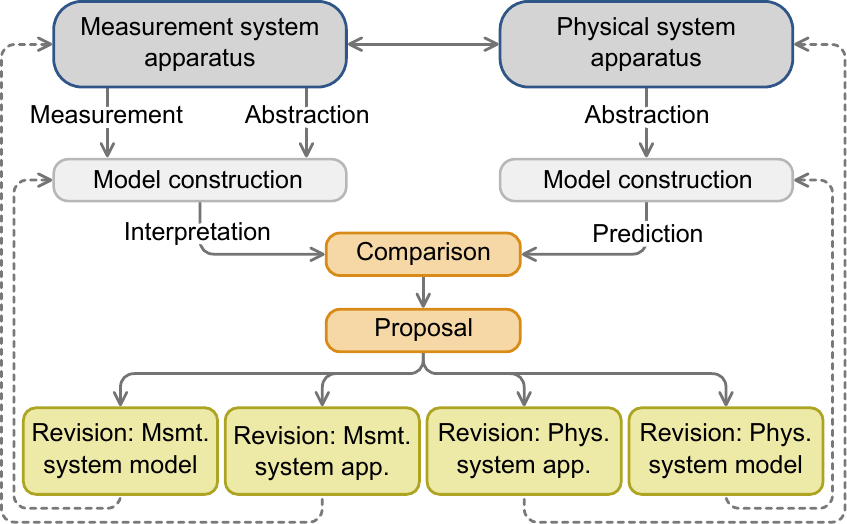}
\caption{\label{fig:framework}Experimental Modeling Framework. The EMF describes the iterative process of modeling the measurement and physical systems, comparing measurements to predictions, proposing explanations for discrepancies, and revising models and/or apparatuses. For brevity, this figure is a simplified version of the full EMF presented by Zwickl et al. in Ref.~\cite{Zwickl2015arXiv}.}
\end{figure}
}

\newcommand{\schematic}{
\begin{figure}[t]
\includegraphics[width=\columnwidth]{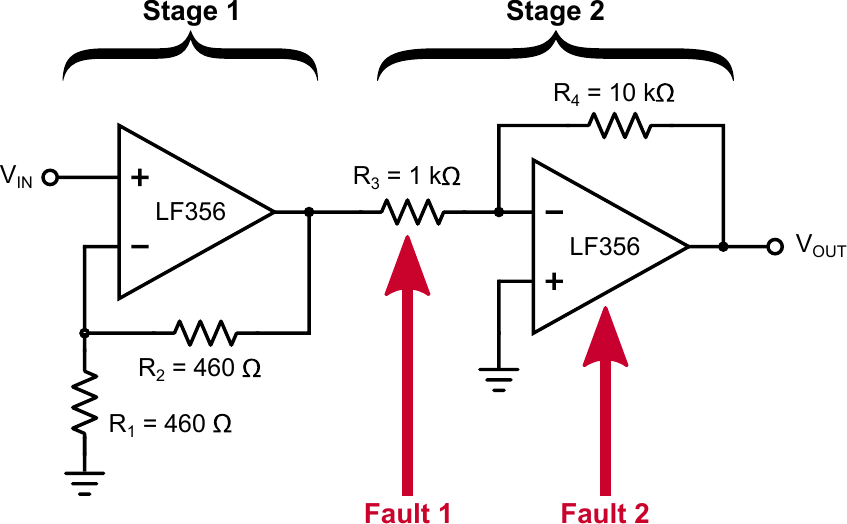}
\caption{\label{fig:diagram}Inverting cascade amplifier. Noninverting (Stage~1) and inverting (Stage~2) amplifiers were connected in series. Two faults were present in Stage 2: the resistor in position $R_3$ had a resistance of 100~\ohm\ rather than 1~k\ohm, and the op-amp was burned out in such a way that its output voltage was equal to the supply voltage regardless of the input voltage.}
\end{figure}
}

\begin{document}

\title{The role of modeling in troubleshooting: an example from electronics}

\author{Dimitri R. Dounas-Frazer}
\affiliation{Department of Physics, University of Colorado Boulder, Boulder, CO 80309, USA}

\author{Kevin L. Van De Bogart}
\affiliation{Department of Physics and Astronomy, University of Maine, Orono, ME 04469, USA}

\author{MacKenzie R. Stetzer}
\affiliation{Department of Physics and Astronomy, University of Maine, Orono, ME 04469, USA}

\author{H. J. Lewandowski}
\affiliation{Department of Physics, University of Colorado Boulder, Boulder, CO 80309, USA}
\affiliation{JILA, National Institute of Standards and Technology and University of Colorado Boulder, Boulder, CO 80309, USA}



\begin{abstract}
Troubleshooting systems is integral to experimental physics in both research and instructional laboratory settings. The recently adopted AAPT Lab Guidelines identify troubleshooting as an important learning outcome of the undergraduate laboratory curriculum. We investigate students' model-based reasoning on a troubleshooting task using data collected in think-aloud interviews during which pairs of students attempted to diagnose and repair a malfunctioning circuit. Our analysis scheme is informed by the Experimental Modeling Framework, which describes physicists' use of mathematical and conceptual models when reasoning about experimental systems. We show that this framework is a useful lens through which to characterize the troubleshooting process.
\end{abstract}

\maketitle

\section{Introduction}

Instructional physics laboratories are well-situated to promote student competence in physics practices, yet few studies focus on such environments---especially at the upper-division level~\cite{DBER2012}. Recently, the American Association of Physics Teachers (AAPT) identified modeling as a major focus area for learning outcomes in undergraduate laboratory courses~\cite{AAPT2015}, consistent with model-based efforts in introductory instruction~\cite{Brewe2008,Etkina2007}. Over the last few years, we have developed the Experimental Modeling Framework~(EMF) to inform curricular transformation of upper-division laboratory courses~\cite{Zwickl2014,*Zwickl2013AJP,*Zwickl2012AIP}. The EMF describes the process through which physicists iteratively revise system models, physical apparatus, and measurement tools in response to discrepant measurements and predictions. The growing emphasis on modeling warrants continued application of the EMF to a broad range of upper-division instructional laboratory contexts.

Previous work demonstrated the utility of the EMF as a tool for characterizing students' model-based reasoning, focusing on an optical physics context~\cite{Zwickl2015arXiv}. In the present work, we use the EMF in a similar capacity but a new context: troubleshooting a malfunctioning electronic circuit. Here ``troubleshooting" refers to the iterative process through which one repairs a malfunctioning apparatus~\cite{Jonassen2006,Schaafstal2000}, a common experimental physics practice. Electronics courses are an ideal environment for studying troubleshooting, due in part to the simplicity of the models and systems with which students interact and the ease with which components can be replaced. Our focus on troubleshooting aligns with the ongoing transformation of an electronics course at the University of Colorado Boulder, informed by both national~\cite{AAPT2015}  and local~\cite{Zwickl2014,*Zwickl2013AJP,*Zwickl2012AIP} learning goals for upper-division laboratory courses.


Of course, the EMF is not the only useful lens through which to understand troubleshooting. In a companion article~\cite{VanDeBogart2015PERC}, we analyze the same data presented here through a different, complementary lens: the Socially Mediated Metacognition Framework (SMMF)~\cite{Goos2002}.  Together, the EMF and SMMF allow us to more fully explore students' strategic navigation of an electronics task. In the present work, we briefly highlight one area where a metacognitive perspective would provide additional insight into students' troubleshooting process. However, our main focus is on application of the EMF to an electronics troubleshooting context.

\vspace{-10pt}

\section{Modeling \& Troubleshooting}

In the the EMF, ``models" are defined as abstract representations used to explain aspects of the real world and predict scientific phenomena~\cite{Zwickl2013AJP,*Zwickl2012AIP}. Models are embedded in known principles and concepts, but contain simplifications and assumptions that yield tractable mathematical, graphical, and other representations. The EMF divides systems into two parts: the \emph{Physical System} and the \emph{Measurement System}. 
In the analysis herein, we focus on student engagement with only the \emph{Physical System}. Accordingly, we do not attend to the \emph{Measurement System} further. A discussion of the interaction between \emph{Physical} and \emph{Measurement Systems} can be found in Ref.~\cite{Zwickl2015arXiv}.

The process of ``modeling" is dynamic and iterative, involving the following phases: \emph{Model Construction}, the process of developing models of the system; \emph{Prediction}, the use of a model to inform expectations about measurements; \emph{Comparison}, the act of comparing measurements to predictions; \emph{Proposal}, the act of proposing a potential explanation for, and/or solution to, discrepant measurements and predictions; and \emph{Revision}, the process of making changes to either the apparatus or model in order to bring measurements and predictions into better alignment. We operationalize \emph{Comparison} as utterances in service of answering the question, ``Is the agreement between measurement and prediction good enough?" Similarly, \emph{Proposal} includes statements that seek to answer the question, ``How can we get better agreement?" A visualization of the modeling process is presented in Fig.~\ref{fig:framework}.

\framework

Similar to modeling, troubleshooting is also a dynamic and iterative process. Troubleshooting can be subdivided into four tasks~\cite{Jonassen2006, Schaafstal2000}: constructing the problem space, identifying fault systems, diagnosing faults, and generating and verifying solutions. These tasks are similar to the modeling phases. For example, construction of the problem space and identification of faults involve both \emph{Model Construction} and \emph{Comparison}. Because troubleshooting involves fixing a malfunctioning apparatus, \emph{Comparison} typically involves comparing measurements of the actual, malfunctioning apparatus to expectations informed by models of an idealized, functioning apparatus. Diagnosis of faults and generation of solutions require \emph{Proposal} and \emph{Revision}. Because the goal of troubleshooting is to repair a malfunctioning system, \emph{Revision} is typically limited to making changes to the apparatus to bring its performance into better alignment with the idealized model.

Troubleshooting further requires a strategic approach. Two examples of strategies present in our data include~\cite{Jonassen2006}: \emph{Trial and Error}, characterized by students arbitrarily focusing on any subsystem in which a fault may be present; and \emph{Split-Half}, characterized by students splitting the system into two subsystems and testing the midpoint in an attempt to isolate the fault in one of the two halves. In this work, we use the EMF as a tool for understanding troubleshooting tasks and strategies.

\vspace{-10pt}

\section{Study Design \& Methods}

\schematic

To facilitate study of both the iterative and strategic aspects of troubleshooting an electronic circuit, we designed the inverting cascade amplifier shown in Fig.~\ref{fig:diagram}. The cascade amplifier consists of two subsystems, or stages: a noninverting amplifer (Stage~1), and an inverting amplifier (Stage~2). Each stage amplifies its input voltage by a multiplicative factor called the gain, $G$, which is determined by the resistor values. In a functional cascade amplifier, Stage~1 would double the input voltage ($G_1=2$) and Stage~2 would both invert the input voltage and amplify it by a factor of ten ($G_2=-10$). In the context of alternating current (ac) signals, ``inverting" is equivalent to shifting the phase of the signal by 180$^\circ$. Because the two stages are connected in series, the overall gain of the cascade amplifier is the product of the gains of Stages 1 and 2: $G_{\mathrm{tot}}= G_1G_2=-20$.

To ensure that students engaged in more than one iteration of the troubleshooting tasks, we introduced two faults in the circuit (Fig.~\ref{fig:diagram}). First, the resistor $R_3$ had a value of 100~\ohm\ rather than the nominal value of 1~k\ohm, increasing the actual gain of Stage~2 by an order of magnitude compared to the nominal gain. Second, we used a broken op-amp in Stage 2, which manifested in a direct current (dc) output voltage of $-15$~V regardless of input. Both faults were localized in Stage~2 so that the cascade amplifier consisted of both a functional subsystem (Stage~1) and a malfunctioning one (Stage~2), making it possible for students to use the \emph{Split-Half} strategy early on in their troubleshooting process.

We conducted think-aloud interviews in which pairs of students were asked to diagnose and repair a faulty inverting cascade amplifier. Students were given a schematic of the circuit, including an algebraic expression for the gain of the entire circuit, but not for either of the two stages. We interviewed 4 pairs of physics majors from each of two institutions (total of 16 students).
Participants were paid volunteers enrolled in junior-level electronics courses at their respective institutions during Fall 2014. Interviews were conducted at the end of the fall term through the beginning of the following spring.

Interviews, which lasted 25--60~minutes, were videotaped and transcribed. We used Interaction Analysis~\cite{Jordan1995} to investigate students' interactions with each other and the apparatus, using the EMF as an \emph{a priori} analysis scheme. We treated models and model-based reasoning as belonging to the shared interactional space; hence we do not expect our results to be a perfect reflection of individual student reasoning.

Two of the authors (D.R.D.F. and K.L.V.D.B.) collaboratively viewed and discussed entire interviews. Select excerpts were interpreted by the research team as a whole. To ensure that our interpretations aligned with those of the broader community, some video excerpts were discussed with research groups and individual researchers outside of our research team.

\vspace{-10pt}

\section{Results \& discussion}\label{sec:Results}

We focus on a 90-second excerpt from one interview. This excerpt was chosen for two reasons: first, it demonstrates connections between \emph{Model Construction}, \emph{Prediction}, \emph{Comparison}, and \emph{Proposal} during students' model-based reasoning; and second, it provides an example of students engaging in socially mediated metacognition, which we discuss elsewhere~\cite{VanDeBogart2015PERC}.

In this excerpt, a pair of students, S1 and S2, troubleshoot the malfunctioning cascade amplifier~(Fig.~\ref{fig:diagram}). The episode takes place about halfway through the task. Prior to this episode, the students computed the expected output of the circuit: given their ac input signal of amplitude 100~mV, the students articulated their expectation that the output of Stage~2 should be an inverted ac signal with amplitude 2~V. Due to the details of their measurement apparatus (the students chose a setting that filters out dc signals), the students measured the output of Stage~2 to be a noninverted ac signal with a 10-mV amplitude. In response to this discrepancy, the students decided to measure the output of Stage~1. This decision is consistent with the \emph{Split-Half} troubleshooting strategy.


The only expectation the students previously articulated about the Stage~1 output is that it should be different from ground. After articulating this expectation, the students measured the output of Stage~1 to be a noninverted ac signal of amplitude 200~mV. The transcript begins immediately after this measurement: \\ \\
\noindent
\begin{tabular}{r>{\bfseries}lp{7.5cm}}
1&S1: &  So, it's doubled V-in, but it's not inverted it.\\
2& &  And it shouldn't be inver-- should be--\\
3&S2: &  Is that an inverting amplifier?\\
4&S1: &  No, it's not.\\
5& & An inverting amplifier is connected to the--\\
6& & V-in is connected to the negative terminal, right?\\
7&S2: &  Yeah, yeah.\\
8&S1: &  So it shouldn't be inverted.
\end{tabular}\\ \\
\noindent
Here S1 stated that the effect of Stage 1 was to double the input voltage but to leave its phase unchanged (Line~1). He then attempted to reconcile his findings with his expectations. However, even though he started to make a prediction, he interrupted himself and did not complete his statement (Line~2). S2 then immediately interrupted S1: to ask whether Stage~1 is an inverting amplifier (Line~3). S1 responded by claiming that the first stage is not an inverting amplifier (Line~4) and justifying this claim by describing a characteristic feature of inverting amplifiers (Lines~5--6). S1 then asked for, and received, confirmation from S2 about his line of reasoning (Lines~6--7). After receiving this affirmation, S1 articulated his expectation about the phase of the Stage~1 output: ``So it [the output of Stage~1] shouldn't be inverted" (Line~8). This utterance is an example of both \emph{Prediction} and \emph{Comparison}: S1 articulated a model-informed expectation about the phase of the Stage~1 output in the context of determining whether the observed output was what the students ``should" be seeing.

The students then went on to discuss the second stage:\\[6pt]
\noindent
\begin{tabular}{r>{\bfseries}lp{7.5cm}}
9&S1:& So this one-- \emph{(Points to schematic)}\\
10&S2: &  Well, neither of them are inverting. \\
11& & Oh, yes. \emph{(Points to schematic)}\\
12 & & This one is inverting.\\
13&S1: &  The second one is inverting.
\end{tabular}\\[6pt]
In Lines~9 and 12, ``this one" is the second stage, as evidenced by Line~13. In this exchange, S1 and S2 correctly identified Stage~2 as an inverting amplifier. 

Together, Lines~3--13 provide an example of \emph{Model Construction} through which the students constructed a model of the inverting cascade amplifier as comprising two distinct amplifier subsystems, one noninverting and the other inverting. The students previously had a model of the circuit that was sufficient to predict the amplitude and phase of the output of the entire circuit, but not of the midpoint. While this preliminary model was made more sophisticated as a result of the \emph{Model Construction} in Lines~3--13, this is not an example of \emph{Revision} because the changes were not made in response to a discrepancy between a prediction and a measurement. In fact, we argue that the lack of a sufficiently sophisticated model prevented S1 from articulating a prediction in Line~2. Only after constructing a model of the circuit as being comprised of a noninverting subsystem (Lines~3--7) was S1 able to make a prediction about the expected output of Stage~1 (Line~8), thus facilitating the \emph{Split-Half} strategy. This connection between \emph{Model Construction} and \emph{Prediction} is consistent with the EMF.


The students then focused the output of Stage~2:\\[6pt]
\noindent
\begin{tabular}{r>{\bfseries}lp{7.5cm}}
14&S2: &  But our V-out right now isn't inverting.\\ 
 & & \emph{(Points to oscilloscope)} \\
15 & & So that probably means that these positive-- plus \\
16 & & and minus terminals on the second one are just \\
17 & & mixed up. \emph{(Points to schematic)}\\
\end{tabular}\\[6pt]
Line~14 is an example of \emph{Comparison}: S2 compared the previously measured output of the circuit to his expectation that the output should be inverted. A prediction is not explicitly articulated in the excerpt; however, one was articulated earlier in the interview. Lines~15--17, on the other hand, are an example of \emph{Proposal}: to explain the discrepancy, S2 proposed that the input terminals of the op-amp in Stage 2 were connected incorrectly. The flow from \emph{Comparison} to \emph{Proposal} is also consistent with the EMF.

The students then discussed S2's proposal further:\\[6pt]
\noindent
\begin{tabular}{r>{\bfseries}lp{7.5cm}}
18&S1: &  Why?\\
19&S2: &  Because it's not inverting.\\
20 & & So this is an inverting amplifier so they just mixed \\
21 & & up the plus and minus. \emph{(Points to schematic)}\\
22&S1: &  But this one's not doing anything at all. \\
& & \emph{(Points to schematic)} \\
23& &The way this is drawn here is inverting.\\
24&S2: &  Yeah. But on here--\emph{(Points to circuit)}\\
25&S1: & On here it's not--  \emph{(Leans over circuit)} \\
26& & There's no output at all. \\
27 & & I mean there's this tiny-- \emph{(Points to oscilloscope)}\\
28&S2: &  What do you mean?  \emph{(Points to oscilloscope)}\\
29 & & Yeah, there's--\\
30&S1: &  I guess, but that's like--\\
31&S2: &  How much-- Well, how big is it?\\
32&S1: &  It's tiny. It's like ten millivolts.\\
33&S2: &  Oh. Well, okay.
\end{tabular}\\[6pt]
In Lines~18--33, the students engaged in a cyclic interaction as a mutual attempt to understand one another's thinking. S2 identified that he was concerned with explaining why the output was not inverted (Line~19). S1 followed by offering his own interpretation of the problem: the output of Stage~2 was negligibly small (Lines~26--32). Each student asked the other to clarify their reasoning (Lines~18 and 28). Ultimately, S2 accepted S1's reasoning~(Line~33).

Student interactions like those presented in Lines~18--33 are common in our data. However, such interactions are not well-captured by the EMF because they do not constitute a modeling phase, but instead represent students' negotiations about whether and when to move from one phase to the next. In this case, the negotiation culminated in a decision \emph{not} to follow up on S2's idea about ``mixed up" inputs (Lines~15--17). A complementary framework to the EMF is needed to fully understand this interaction. In a companion study, we investigate this interaction from a metacognitive perspective~\cite{VanDeBogart2015PERC}.

S2 then stepped back to consolidate his thoughts:\\[6pt]
\noindent
\begin{tabular}{r>{\bfseries}lp{7.5cm}}
34& S2: & We have a good output for the first op-amp,\\
35& &  so we are going to have--\\
36& & the problem is in the second one.
\end{tabular}\\ \\
Line~34 is another example of \emph{Comparison}: S2 reiterated that, regarding the output of Stage~1, the students' measurements and expectations were in good enough agreement with one another. In this case, ``good enough" meant ``good enough to eliminate Stage 1 as a potential source of fault." While there is no explicit comparison being made in Lines~35--36, we nevertheless infer that S2 made an appraisal for the second stage as well. The lack of good enough agreement between measurements and expectations in Stage~2 led S2 to conclude that ``the problem is in the second one" (Line~36). From a troubleshooting perspective, Line~36 represents the culmination of the \emph{Split-Half} strategy, which enabled the students to narrow their search for faults to a single subsystem of the cascade amplifier.
\vspace{-4pt}

\section{Summary \& Future Directions}
We applied the EMF to an activity in which a pair of students troubleshot a malfunctioning electronic circuit, demonstrating that students engage in model-based reasoning throughout the troubleshooting process. In particular, we discussed the interaction of various EMF phases during a troubleshooting cycle in which students employed the \emph{Split-Half} strategy. We further identified an example of student reasoning that is beyond the scope of the EMF: collaborative decisions about how to navigate between phases. Elsewhere, we discuss a metacognitive framework for describing this process~\cite{VanDeBogart2015PERC}.

In addition to its utility in characterizing model-based reasoning, the EMF is a useful tool for facilitating curricular transformation of laboratory courses. Along these lines, our work will inform explicit instruction and assessment of troubleshooting skills in electronics courses.


\vspace{-6pt}

\acknowledgments The authors acknowledge A. Little, B. Zwickl, B. Wilcox, G. Quan, J. Stanley, and the Colorado and Maryland PER groups for useful input on study design and interpretation of results. This work was supported by NSF grants DUE-1323101, DUE-1323426, DUE-1245313, and DUE-0962805.


%

\end{document}